\documentclass[twocolumn,floats,prl,aps]{revtex4}

\usepackage[dvips]{graphicx}
\usepackage{amsfonts}
\usepackage{amsmath}
\usepackage{amssymb}
\usepackage{exscale}
\usepackage{eufrak}

\begin{document}

\draft    % Show PACS

%\wideabs{    %>>> Not to use in preprint mode.

\title{Comment to the paper {\it ``In-plane spectral weight shift of charge-carriers in 
	   YBa$_{2}$Cu$_{3}$O$_{6.9}$"} by A.~V.~Boris {\it et al.}, 
	   Science \underline{304}, 708 (2004)
	   }
\author{A.F.~Santander-Syro$^{1}$, and N.~Bontemps$^{2}$}

\address{$^{1}$Laboratoire de Physique des Solides, CNRS UMR 8502,
         Universit\'e Paris-Sud, 91405 Orsay cedex, France}
         
\address{$^{2}$Laboratoire de Physique du Solide, CNRS UPR 5,
         Ecole Sup\'erieure de Physique et Chimie
         Industrielles de la Ville de Paris,
         75231 Paris cedex 5, France}

\date{\today}

%\maketitle %>>> To use with standard RevTeX.  With RevTeX4, \maketitle
            %>>> must be placed after the abstract and the pacs.

\begin{abstract}
Recently, A. V. Boris and colleagues~\cite{BorisScience} claimed to deduce a decrease 
of intraband spectral weight (SW), and a transfer of SW from 
intraband to interband frequencies, when optimally-doped or slightly underdoped
cuprates become superconducting. We show that, while their data agree 
with others~\cite{vdM-Science,AFSS-EPL}, their analysis is flawed.
They cannot disprove the results which yield a superconductivity-induced
increase of the intraband spectral 
weight, and a transfer of SW from high to low frequencies, 
in underdoped or nearly optimally doped 
Bi$_2$Sr$_2$CaCu$_2$O$_{8+\delta}$~\cite{vdM-Science,AFSS-EPL}.
\end{abstract}

%\pacs{74.25.-q, 74.25.Gz, 74.72.Hs}

\maketitle  %>>> To use with RevTeX4.  Not to use with standard RevTeX.

%} %Not to use in preprint mode

%%%%%%%%%%%%%%%%%%%%%%%%%%%%%%%%%%%%%%%%%%%%%%%%%%%%%%%%%
Recently, A. V. Boris {\it et al.}~\cite{BorisScience} reported ellipsometric 
measurements of the complex dielectric function 
$\epsilon(\omega)=\epsilon_{1}(\omega)+i\epsilon_{2}(\omega)= %
\epsilon_{\infty}+4\pi i\sigma(\omega)/\omega$ in the spectral
range $\hbar\omega=0.01 - 5.6$~eV
in two cuprate superconductors: 
an optimally doped YBa$_2$Cu$_3$O$_{7-\delta}$ (YBCO)
and a slightly underdoped Bi$_2$Sr$_2$CaCu$_2$O$_{8+\delta}$ (Bi-2212). 
They analyzed the temperature changes of $\epsilon_{1}(\omega)$ and
$\sigma_{1}(\omega)$ and claim to extract eventually the normal (N) 
to superconducting (SC) change of intraband spectral weight. 
What is at stake is how to estimate this latter quantity. Its exact value 
is not accessible experimentally because of the presence, close to the conduction band, 
of interband transitions. The only experimental 
quantity which comes as close as possible is what we define
as the conduction-band spectral weight (CSW)~\cite{vdM-Science,AFSS-EPL}:  
\begin{equation}
	CSW(\Omega,T)=\int_{0}^{\Omega}\sigma_{1}(\omega,T)d\omega,
\label{EqCSW}
\end{equation}
The upper integration limit $\Omega$ is chosen so as to span the excitations within the
conduction band only, yet low enough in order not to include interband transitions. 
A typical value for $\Omega$ is 1~eV~\cite{vdM-Science,AFSS-EPL}. 
In the superconducting state,
this integral contains the $\delta$-function at $\omega=0$ which is the
contribution from the superfluid.

In the spectral range $0.15 - 4$~eV for YBCO, 
and $0.15-0.6$~eV for Bi-2212 (see the supporting on-line material
of Ref.~\cite{BorisScience}),
the data displayed by Boris {\it et al.}
show that: {\it i)} $\Delta\sigma_{1,NS}=\sigma_{1,N}-\sigma_{1,SC}$ is positive, 
{\it ii)} $\Delta\epsilon_{1,NS}=\epsilon_{1,N}-\epsilon_{1,SC}$ 
%crosses zero and 
becomes, within their noise, indistinguishable from zero. 

Their claim is then that one can obtain from Kramers-Kronig relations 
``a conceptually model independent estimate of whether SW is redistributed 
from high to low energy or vice versa". 
The bottom line of their argument can be expressed as follows~\cite{Note1}:
if $\Delta\sigma_{1,NS}$ is positive in a narrow region $\omega_0 \pm \delta \omega$
around $\omega_0$, whereas $\Delta\epsilon_{1,NS}$ is zero in this region,
this positive $\Delta\sigma_{1,NS}$ needs to be balanced by a 
corresponding negative counterpart below and above the narrow region
$\omega_0 \pm \delta \omega$
(supporting on-line material of Ref.~\cite{BorisScience}).
However, one can show that, in order to be correct, 
this argument should be supplemented by an assumption~\cite{Roland}:
the changes $\Delta\sigma_{1,NS}$ below and above the narrow region
$\omega_0 \pm \delta \omega$
keep the same (negative) sign at {\it all} frequencies. 

We provide here firstly a simple counter example 
that contradicts their so-called ``model independent conclusion based 
solely on KK relation" ~\cite{BorisScience}.  
In the normal state, we assume a Drude oscillator 
(plasma frequency $\Omega_N$, width $\Gamma$) 
and an interband transition schematically represented by a $\delta$-function 
(but it could be a Lorentz oscillator as well) centered at a frequency
$\Omega_T>\Omega_N$ beyond the conduction band,
with a weight $\Omega_{B}^{2}$. 
In the SC state, all the SW of the interband transition, and part of the SW of the 
Drude oscillator, are transferred to the $\delta$-function at $\omega=0$ 
(London frequency $\Omega_{L}$).  
The width of the Drude oscillator is left constant for simplicity.  
We chose to define the spectral functions in the $\{0,+\infty\}$ range, 
but they can be extended to $\{-\infty,0\}$ 
by symmetry without altering the result. 
This toy model warrants causality and charge-conservation. 
One can show that there exists a set of parameters satisfying the condition 
(which is not stringent)
\begin{equation}
	1-\frac{\Omega_{T}^{2}}{\omega_{0}^{2}} = \frac{\Omega_{B}^{2}}{\Omega_{L}^{2}}
			\frac{\Omega_{T}^{2}+\Gamma^2}{\Gamma^2} < 1,
\label{EqWo}
\end{equation}
(where $\omega_{0} > \Omega_{T}$) such that $\Delta \epsilon_{1,NS}(\omega_{0})=0$
and $\Delta \sigma_{1,NS}(\omega_{0})>0$, 
thus verifying the initial conditions leading to the reasoning of Boris {\it et al.}
However, by construction, the difference ``N minus SC" in spectral weight 
in this toy model is positive at {\it all} frequencies {\it above} zero
(SW is lost at all finite frequencies), and is negative only at $\omega=0$
(gain of SW only at zero frequency, associated with the onset of $\delta(\omega)$
 due to the superfluid).
Besides, because of the transfer of SW from the interband transition 
to the $\delta(\omega)$  function, the CSW of this toy-model is {\it larger}
in the superconducting state than in the normal state.  Thus, this toy model 
contradicts the conclusions from Boris {\it et al.}, although satifying
the same initial conditions. 
A somewhat more ``realistic" model is
developed in the appendix, using a set of Drude, Lorentz and London oscillators. 
This latter model captures the essential observations of Boris and coworkers (Fig.1), 
notably in their slightly underdoped Bi-2212 sample (supporting 
on-line material of Ref.~\cite{BorisScience}). 
Yet it contradicts again the ``model-independent" conclusions of these authors.
Therefore one cannot make any general statement 
based only on Kramers-Kronig relations. The conclusion of reference~\cite{BorisScience}, 
{\it i.e.} ``there must be an overall decrease of 
SW up to 1.5~eV"~\cite{NoteSW}, 
is incorrect just relying on such considerations and cannot therefore 
contradict references~\cite{vdM-Science,AFSS-EPL}. 

We note next that Boris {\it et al.} never compute explicitly the CSW (eq.~\ref{EqCSW}), 
in contrast with references~\cite{vdM-Science,AFSS-EPL}. 
Instead, they use the ``generalized plasma frequency" from the
``extended Drude formalism" (incidentally, this formalism was shown not to be valid 
in the SC state~\cite{Maksimov-NoDG}):
\begin{equation}
	[\Omega_{pl}^{\star}(\omega)]^{2} = 
	\frac{\omega^2}{\textrm{Re}\{[\epsilon_\infty - \epsilon(\omega)]^{-1}\}}.
\label{EqDG}     
\end{equation}
Boris {\it et al.} then claim that  
$\Delta [\Omega_{pl}^{\star}(\omega)]^{2} = [\Omega_{pl,N}^{\star}(\omega)]^{2} -
[\Omega_{pl,SC}^{\star}(\omega)]^{2}$, {\it i.e.} the change
of the square of the generalized plasma frequency, 
can serve to compute the change in CSW, including its sign. 
The latter statement is incorrect.
Here again, a direct computation (not shown, but easily implemented) 
from our toy-model  yields changes in  
$[\Omega_{pl}^{\star}(\omega)]^{2}$ and in the CSW 
with {\it opposite frequency-dependent trends} and {\it different signs} 
over a sizeable portion of the conduction band.

The difference in {\it sign} is also shown (Fig.2) in the more realistic model 
developed in the appendix.

Finally, note that, while in {\it underdoped} Bi-2212 there is a SC-induced
increase of the CSW~\cite{vdM-Science,AFSS-EPL,AFSS-BigOne}, 
  we observe the different trend in {\it overdoped} Bi-2212 (having $T_{c}=63$~K),
namely a change in CSW that is compatible, in sign and magnitude, with the predictions
of the BCS theory~\cite{GD-AFSS}. We observe that going 
from the overdoped to the underdoped regime, the SC-induced change in CSW 
is actually progressive, going through zero not far from optimum doping. 
This progressive change strongly suggests that there is in the cuprates 
a smooth transition from a conventional mode of condensation in the overdoped regime 
to an unconventional mode in the underdoped one~\cite{GD-AFSS}. 

In conclusion, whereas the data presented by Boris and colleagues~\cite{BorisScience} 
on {\it nearly optimally-doped cuprates} actually {\it agree} with previous work 
in underdoped to optimally doped cuprates, their conclusions are flawed 
by an incorrect or model-dependent data analysis. They cannot disprove work 
based on the integration of $\sigma_{1}(\omega)$ 
(the only correct experimental way to quantify  the CSW change), 
which yields, for underdoped samples, a superconductivity-induced transfer of SW 
{\it from high to low energies} involving an energy scale 
beyond 1~eV~\cite{vdM-Science,AFSS-EPL,AFSS-BigOne}.  

We are grateful to Roland Combescot for very fruitful discussions and critical reading of this comment.

%-----------------------------------------
%Table DLL parameters
%
\begin{table*}
  \begin{center}
   \begin{scriptsize}
    \begin{tabular}{|c|c|c|}
      \hline
      %Line 1      
      {\bf Normal state} & & {\bf Superconducting state} \\
      \hline \hline
      %Line 2
      $\Omega_{P,N}=8200$~cm$^{-1}$; $\Gamma_{P,N}=160$~cm$^{-1}$ & & $\Omega_{P,S}=6200$~cm$^{-1}$; $\Gamma_{P,S}=160$~cm$^{-1}$ \\
      %Line 3
       & & $\Omega_{L}^{2} = 28999702.2$~cm$^{-2}$ \\
      \hline
      %Line 4
      \begin{tabular}{r|r|r}
      $\Omega_{T,j}$ [cm$^{-1}$] & $\Gamma_{T,j}$ [cm$^{-1}$] & $\Delta \epsilon_{j}$ \\
      \hline
      432		& 41		& 0.8 \\
      478		& 63		& 4.4 \\
      528		& 44		& 1.9 \\
      693		& 100		& 0.5 \\
      703		& 66		& 0.5 \\
      765		& 200		& 2.3 \\
      861		& 475		& 5.4 \\
      1105		& 1450		& 32.7 \\
      2060		& 5250		& 21.8 \\
      3685		& 5100		& 7.3 \\
      14000		& 18000		& 0.01 \\
      17000		& 10000		& 0.2 \\
      30000		& 20000		& 0.9 \\
      \end{tabular}
      & &
      \begin{tabular}{r|r|r}
      $\Omega_{T,j}$ [cm$^{-1}$] & $\Gamma_{T,j}$ [cm$^{-1}$] & $\Delta \epsilon_{j}$ \\
      \hline
      --		& --		& -- \\
      477		& 48		& 2.2 \\
      480		& 205		& 2.8 \\
      --		& --		& -- \\
      703		& 600		& 2.3 \\
      --		& --		& -- \\
      870		& 550		& 7.2 \\
      1116		& 1500		& 34.6 \\
      1980		& 5600		& 22.7 \\
      3685		& 5000		& 7.44 \\
      14000		& 18000		& 0 \\
      17000		& 10000		& 0.2 \\
      30000		& 20000		& 0.9 \\
      \end{tabular} \\
      \hline    
    \end{tabular}
   \end{scriptsize}
  \end{center}
  \caption{\footnotesize Parameters for the Drude, Lorentz and London oscillators
  		   used in the discussion following Eq.\ref{Eq:DLL}.  Charge conservation
  		   (Eq.\ref{Eq:ChargeCons}) is ensured by construction.
  		   The optical functions issued from these parameters yield the changes
  		   in $\epsilon_{1}(\omega)$, $\sigma_{1}(\omega)$ and in spectral weight
  		   shown in figures~\ref{Fig1} and \ref{Fig2}.}
  \label{Table1}
\end{table*}
%-------------------------------
%

%%%%%%%%%%%%%%%%%%%%%%%%%%%%%%%%%
\section{Appendix}

%-------------
\begin{figure}
  \begin{center}
    \includegraphics[width=8cm]{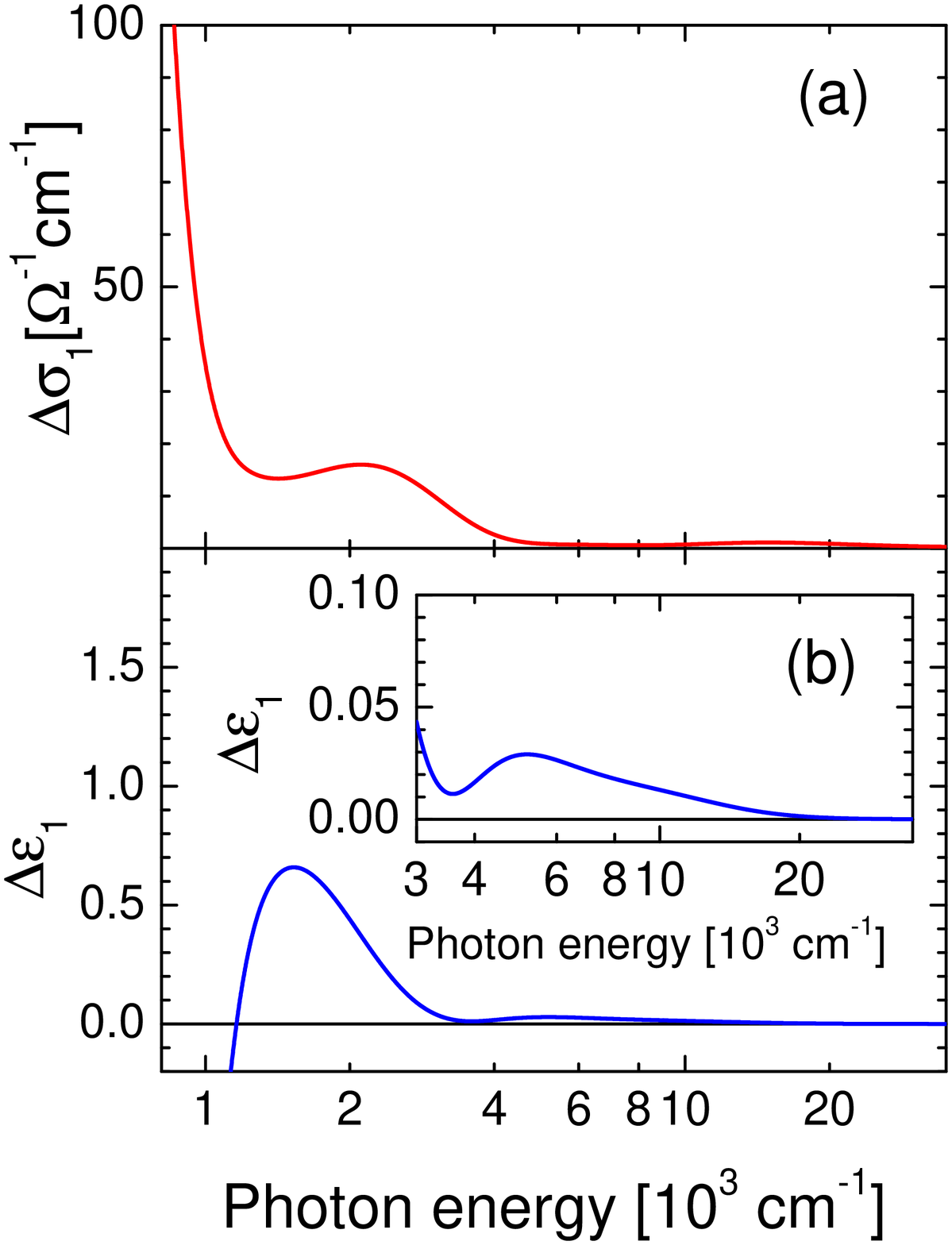}
  \end{center}
  \caption{\label{Fig1}
         Superconductivity-induced changes in the optical functions resulting from the
         model of Eq.~\ref{Eq:DLL} and table~\ref{Table1}: 
         (a) $\Delta\sigma_{1,NS}(\omega)$; 
         (b) $\Delta\epsilon_{1,NS}(\omega)$.  The inset presents a zoom of
         $\Delta\epsilon_{1,NS}(\omega)$ from 0.4 to 4~eV, showing that the
         change in this function is lower than 0.03 above 0.4~eV (and lower
         than 0.01 above 1.45~eV).  The scatter in the data
         of Boris {\it et al.} is about $\pm 0.1$ 
         (see figure~1B in Ref.\cite{BorisScience} and figure~S5 in 
         their supporting on-line material), {\it i.e.}, twice the size of the inset
         shown here.
         }
\end{figure}
%
%-------------
\begin{figure}
  \begin{center}
    \includegraphics[width=8cm]{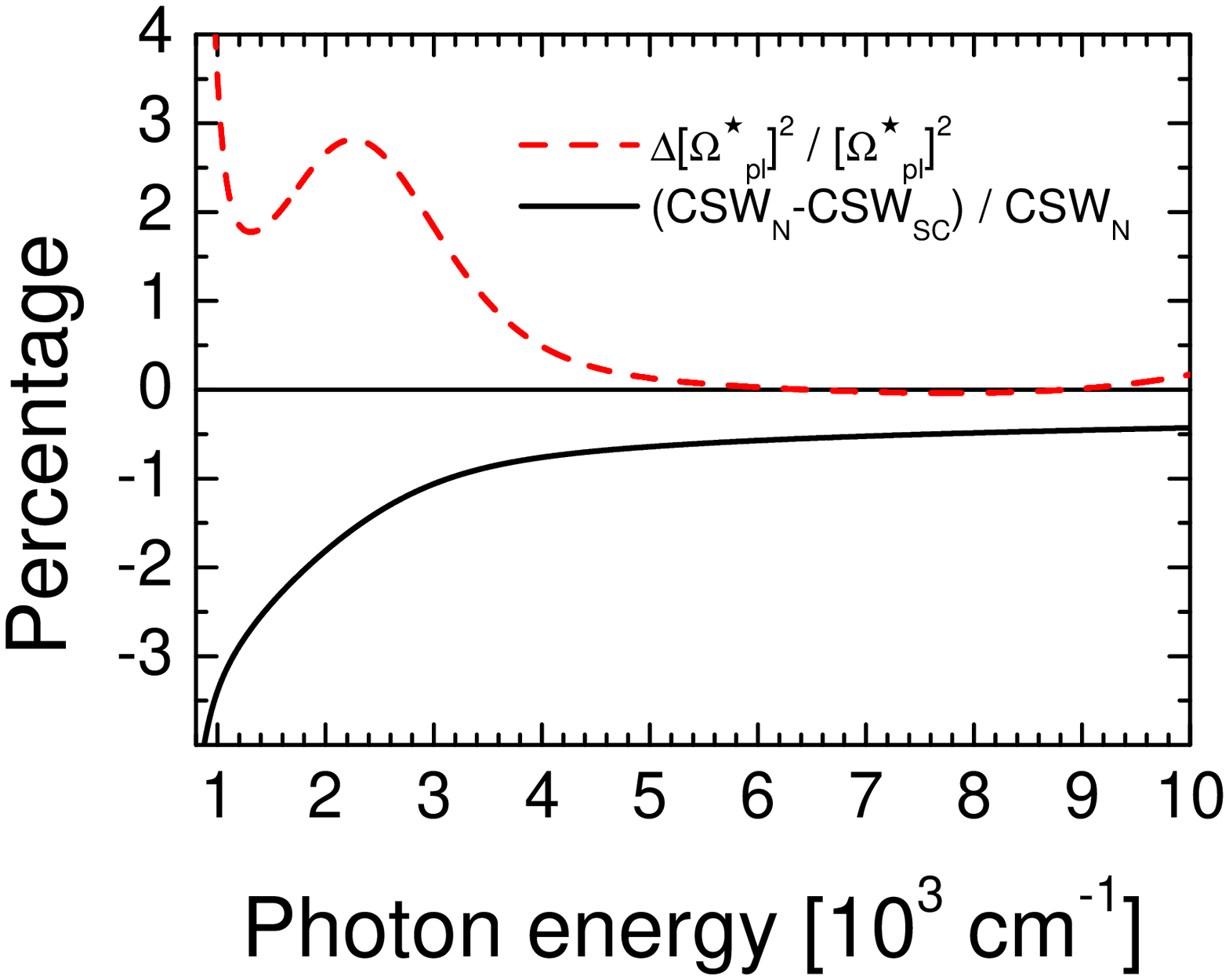}
  \end{center}
  \caption{\label{Fig2}
         Superconductivity-induced change of the conduction-band spectral
         weight for the model of Eq.~\ref{Eq:DLL} and 
         table~\ref{Table1}, as computed from
         equation~\ref{EqCSW} (solid line), and comparison with the
         change of ``generalized spectral weight" (dashed line) 
         calculated from the generalized Drude formula (as done in
         reference~\cite{BorisScience}) on the same model functions.
         Note that, over the whole conduction band, both quantities
         differ by more than 0.5\%, do not track each other 
         and display different signs
         over more than one-half of the normal-state plasma frequency. 
         }
\end{figure}
%-------------

In this section we address the problem of $\Delta \epsilon_{1} (\omega)$
remaining vanishingly small
(within the experimental uncertainty) over a large frequency range, 
which we quote from 
Boris and coworkers~\cite{BorisScience}:  ``...$\epsilon_{1a} (\omega)$ 
remains virtually T-independent. This trend holds not only near but 
persists over a wide energy range 
from 1.5 down to at least 0.15~eV. The KK relation necessarily implies 
that the SW loss between 
0.15 and 1.5~eV needs to be balanced by a corresponding SW gain below 
0.15~eV and above 1.5~eV."

We now use an ensemble of Drude, Lorentz and London oscillators to write the
optical functions of our model system in the N and SC states:
\begin{eqnarray}
  \epsilon(\omega) & = & 1 - %
  \frac{\Omega_{P}^{2}}{\omega (\omega + i \Gamma_{P})} +
  \nonumber \\ %
  & & + \sum_{j} \frac{\Delta \epsilon_{j} \Omega_{T,j}^{2}} %
  {(\Omega_{T,j}^{2}-\omega^{2}) - i \Gamma_{T,j} \omega} + 
  \nonumber \\%
  & & + \chi_{\textsc{London}}(\omega),
\label{Eq:DLL}
\end{eqnarray}
where $(\Omega_{P},\Gamma_{P})$ are the plasma frequency and scattering-rate,
$\Delta \epsilon_{j}$ is the oscillator strength of the oscillator 
centered at $\Omega_{T,j}$ with damping $\Gamma_{T,j}$, and
where the London susceptibility
\begin{equation}
	\chi_{\textsc{London}}(\omega)=-\frac{\Omega_{L}^{2}}{\omega^{2}} %
    + i \frac{\pi \Omega_{L}^{2}}{2 \omega} \delta(\omega)
\label{Eq:XiLondon}
\end{equation}
is used only in the SC state. 
For simplicity, we set $\epsilon_{\infty}=1$ in both the N and SC states. 
We choose the ensemble of parameters
displayed in table~\ref{Table1}.  These parameters are such that,
\begin{equation}
	\Omega_{P}^{2}+\Omega_{L}^{2}+\sum_{j}\Delta\epsilon_{j}\Omega_{T,j}^{2}=
	\textrm{const.},
\label{Eq:ChargeCons}
\end{equation}
{\it i.e.}, the spectral weight is exactly the same in the N and SC states
when integrating from zero to infinity.  We have set
the N and SC plasma scattering rates {\it identical}, so as to simulate a N-to-SC
transition %at the same ($T=0$) temperature 
 without any temperature-dependent
narrowing of the Drude term, thus ruling out by construction any SW transfer 
towards low energy due to narrowing of the Drude contribution.

This model captures the essential observations 
of Boris and coworkers in their slightly underdoped Bi-2212 sample, 
namely $\Delta\sigma_{1,NS}(\omega)>0$ for all $\omega>0$,
and $\Delta\epsilon_{1,NS}(\omega) \approx 0$ (within $0.03$ or lower) beyond
$\omega \approx 0.4$~eV (compare Fig.~1 in this comment to figure S5 from the supporting 
on-line material of \cite{BorisScience}). 
Note that, 
while $\Delta\sigma_{1,NS}(\omega)$ remains positive for all $\omega>0$,
one can always choose an arbitrarly wide energy range in which
the two conditions for the reasoning of Boris and coworkers are fulfilled
within any experimental resolution (in our model, $\Delta\epsilon_{1,NS}(\omega)$ 
remains bounded between $+0.03$ and $-2 \times 10^{-16}$).
Yet, despite the qualitative and quantitative similarity, 
the conclusions of Boris {\it et al.} do not
hold  in this model.  
SW is transferred only {\it from high to low frequencies} (including
the SC condensate), so that there is a SC-induced {\it increase} of
the conduction-band spectral weight (the difference ``N minus SC" is negative), 
as shown in Fig.~2.  This
figure shows as well that the the change in $[\Omega_{pl}^{\star}(\omega)]^{2}$
does not allow to track either the sign or the magnitude of the CSW change, so
that the extended Drude formalism is not useful in this case.  We have found
as well other sets of parameters that reconfirm all our conclusions.  
Some of these sets even yield a positive sign for the change of the generalized
plasma frequency over the full conduction band, whereas a negative sign is observed
in the change of the conduction-band spectral weight.

%%%%%%%%%%%%%%%%%%%%%%%%%%%%%%%%%%%%%%%%%%%%%%%%%%%%%%%%%

%%%%%%%%%%%%%%%%%%%%%%%%%%%%%%%%%%%%%%%%%%%%%%%%%%%%%%%%%
\end{document}